\documentclass[default]{sn-jnl}

\usepackage{amsmath,amsfonts,amssymb}
\usepackage{graphicx}
\usepackage{adjustbox}
\usepackage{graphicx}
\usepackage{siunitx}
\usepackage{mathtools}
\usepackage{threeparttable}
\usepackage{float}
\usepackage{booktabs}
\usepackage{tabularx}
\usepackage{makecell}
\newcolumntype{C}[1]{>{\centering\arraybackslash}p{#1}}
\newcolumntype{L}[1]{>{\raggedright\arraybackslash}p{#1}}
\usepackage{multicol}
\usepackage[T1]{fontenc}
\usepackage{csquotes}
\usepackage{url}
\usepackage{natbib}
\usepackage{caption}
\usepackage{subcaption}

\graphicspath{ 
{fig_results/}
}

\unnumbered

\begin{document}

\title[Universal Lymph Node Detection in T2 MRI using Neural Networks]{Universal Lymph Node Detection in T2 MRI using Neural Networks}

\author*[1]{\fnm{Tejas Sudharshan} \sur{Mathai}}
\author[1]{\fnm{Sungwon} \sur{Lee}}
\author[1]{\fnm{Thomas C.} \sur{Shen}}
\author[2]{\fnm{Zhiyong} \sur{Lu}}
\author[1]{\fnm{Ronald M.} \sur{Summers}}


\affil[1]{\orgdiv{Imaging Biomarkers and Computer-Aided Diagnosis Laboratory, Clinical Center}, \orgname{NIH}, \orgaddress{\city{Bethesda}, \state{MD}, \country{USA}}}

\affil[2]{\orgdiv{National Center for Biotechnology Information, National Library of Medicine}, \orgname{NIH}, \orgaddress{\city{Bethesda}, \state{MD}, \country{USA}}}


\abstract{

\textbf{Purpose:} Identification of abdominal Lymph Nodes (LN) that are suspicious for metastasis in T2 Magnetic Resonance Imaging (MRI) scans is critical for staging of lymphoproliferative diseases. Prior work on LN detection has been limited to specific anatomical regions of the body (pelvis, rectum) in single MR slices. Therefore, the development of a universal approach to detect LN in full T2 MRI volumes is highly desirable. 

\textbf{Methods:} In this study, a Computer Aided Detection (CAD) pipeline to universally identify abdominal LN in volumetric T2 MRI using neural networks is proposed. First, we trained various neural network models for detecting LN: Faster RCNN with and without Hard Negative Example Mining (HNEM), FCOS, FoveaBox, VFNet, and Detection Transformer (DETR). Next, we show that the state-of-the-art (SOTA) VFNet model with Adaptive Training Sample Selection (ATSS) outperforms Faster RCNN with HNEM. Finally, we ensembled models that surpassed a 45\% mAP threshold. We found that the VFNet model and one-stage model ensemble can be interchangeably used in the CAD pipeline. 

\textbf{Results:} Experiments on 122 test T2 MRI volumes revealed that VFNet achieved a 51.1\% mAP and 78.7\% recall at 4 false positives (FP) per volume, while the one-stage model ensemble achieved a mAP of 52.3\% and sensitivity of 78.7\% at 4FP. 

\textbf{Conclusion:} Our contribution is a CAD pipeline that detects LN in T2 MRI volumes, resulting in a sensitivity improvement of $\sim$14 points over the current SOTA method for LN detection (sensitivity of 78.7\% at 4 FP vs. 64.6\% at 5 FP per volume). 

}


\maketitle


\section{Introduction}
\label{sec_Introduction}

The lymphatic system consists of lymph nodes that fight infection in the body by ridding it of foreign matter. In lympho-proliferative diseases, patients may present with enlarged and metastatic LN, which need to be identified and distinguished from non-metastatic LN. Furthermore, cancerous LN appearing at sites in the body that do not correlate with the first site of lymphatic spread indicate distant metastasis \cite{Amin2017}. LN are visualized using multi-parametric MRI, e.g. T2-weighted MRI, and radiologists use the AJCC guidelines \cite{Amin2017} to measure the size of a LN through its long and short axis diameters (LAD and SAD). Nodal size predominantly defines the cancer management and therapy; nodes with a SAD of $\geq$ 10mm are considered suspicious for cancer \cite{Taupitz2007}. It is often difficult for a radiologist to identify and size LN due to their inconsistent shapes, varying appearances, and diverse anatomical locations. Typically, they utilize another MRI scan, such as Diffusion Weighted Imaging (DWI), to confirm their findings. Usually, multiple scanners with different exam protocols are used in a clinical setting to image patients. Despite the complimentary sources of information, radiologists can sometimes miss suspicious LN. Due to these challenging circumstances involving imaging and radiology workflow issues, an automated CAD pipeline is needed for identification of LN in T2 MRI scans such that LN sizing can be completed. 

There has been a moderate amount of prior work \cite{Zhao2020_mri,Debats2019_mri,Lu2018_mri,Wang2022_mri,Mathai2021_mlmi,Mathai2021_spie} on MRI-based LN detection, but they focus on specific anatomical regions. A Mask-RCNN network was used in \cite{Zhao2020_mri} to identify and segment LN in pelvic MRI images with 62.6\% recall at 8.2 FP per volume. In \cite{Debats2019_mri}, a GentleBoost classifier followed by a convolutional neural network (CNN) was used to achieve a 85\% LN detection sensitivity at 5-10 FP per image. In \cite{Lu2018_mri}, Faster-RCNN identified pelvic LN with an AUC of 91.2\%. Universal LN detection in the abdomen has only been researched recently with \cite{Wang2022_mri} describing a 64.6\% sensitivity at 5 FP per volume, while \cite{Mathai2021_mlmi} posted a 91.6\% sensitivity at 4 FP per image on 2D slices only. 

In this study, we design a CAD pipeline that detects LN in T2-weighted MRI volumes. We evaluate SOTA off-the-shelf detection models, such as VFNet \cite{Zhang2021_vfnet}, DETR transformer \cite{Carion2020_detr}, FoveaBox \cite{Kong2019_foveabox}, Faster RCNN \cite{Ren2015_fasterrcnn} etc., on the LN detection task. Next, we describe the superiority of the VFNet model with Adaptive Training Sample Selection (ATSS) \cite{Zhang2021_vfnet} over a Faster RCNN model trained with HNEM \cite{Tang2019_ULDOR}. Finally, models that exceed a 45\% mAP threshold are ensembled together, and we observe that the one-stage model ensemble and the VFNet model are similarly proficient at detecting LN in T2 MRI scans: a $\geq$51\% mean average precision (mAP) and $\sim$78.7\% sensitivity at 4FP per volume. We do not use DWI scans so that we can compare our results against prior work (although the addition of DWI would enable reliable LN detection). Despite this, an improvement in sensitivity by $\sim$14 points (78.7\% recall at 4FP vs. 64.6\% recall at 5P per volume) is achieved over the most recent SOTA LN detection method \cite{Wang2022_mri}. 

\begin{figure}[!ht]
\centering
\begin{subfigure}[b]{0.24\columnwidth}
\vspace*{\fill}
  \centering
  \includegraphics[width=\columnwidth,height=2.6cm]{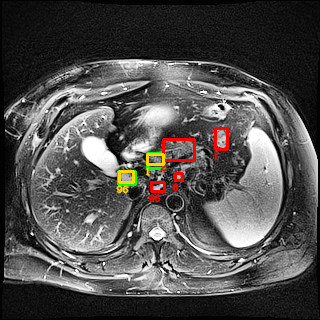}
  \centerline{(a) F-RCNN} 
  \includegraphics[width=\columnwidth,height=2.6cm]{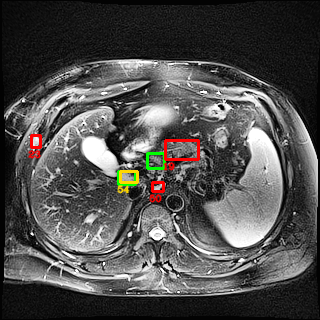}
  \centerline{(e) FoveaBox} 
\end{subfigure} 
\begin{subfigure}[b]{0.24\columnwidth}
\vspace*{\fill}
  \centering
  \includegraphics[width=\columnwidth,height=2.6cm]{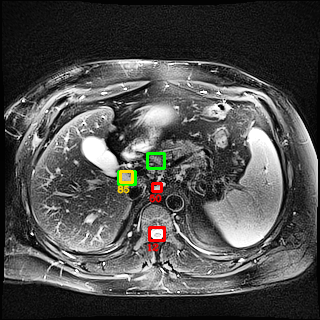}
  \centerline{(b) F-RCNN + HN} 
  \includegraphics[width=\columnwidth,height=2.6cm]{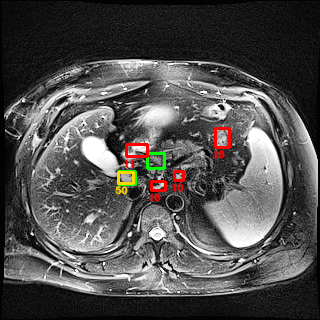}
  \centerline{(f) DETR} 
\end{subfigure} 
\begin{subfigure}[b]{0.24\columnwidth}
\vspace*{\fill}
  \centering
  \includegraphics[width=\columnwidth,height=2.6cm]{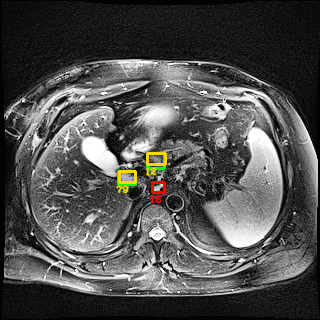}
  \centerline{(c) VFNet} 
  \includegraphics[width=\columnwidth,height=2.6cm]{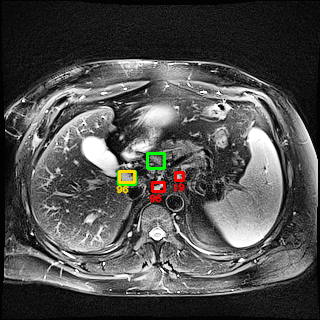}
  \centerline{(g) Ens All} 
\end{subfigure} 
\begin{subfigure}[b]{0.24\columnwidth}
\vspace*{\fill}
  \centering
  \includegraphics[width=\columnwidth,height=2.6cm]{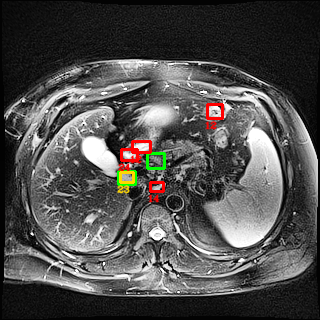}
  \centerline{(d) FCOS}
  \includegraphics[width=\columnwidth,height=2.6cm]{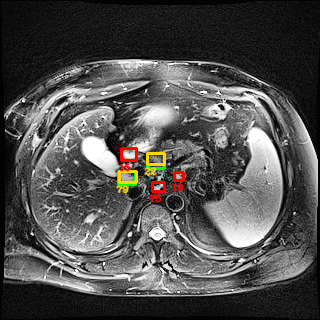}
  \captionsetup{font={footnotesize}}
  \centerline{(h) Ens OS} 
\end{subfigure} 
\caption{LN detection results by different neural networks on a slice from a T2 MRI volume. Green boxes: ground truth, yellow: true positives, and red: false positives. (a) and (b) show the Faster RCNN model without and with HNEM respectively. Notice the reduced FP count in (b). In (c) and (d), LN detection results by VFNet with ATSS and FCOS is shown. More FP are detected by FCOS. (e) and (f) show the results from other SOTA detectors such as FoveaBox and DETR. Finally, (g) shows the ensemble detection result using all models (Faster RCNN + HNEM, VFNet, FoveaBox and DETR), while (h) shows the result of using only the one-stage (OS) model ensemble (VFNet, FoveaBox and DETR). Note the missed LN in (g) captured in (h).}
\label{fig:money}
\end{figure}

\section{Methods}
\label{sec_Methods}

\noindent
\textbf{Object Detectors.} The following were the object detection models that were used in this paper: 1) Faster R-CNN \cite{Ren2015_fasterrcnn}, 2) VFNet \cite{Zhang2021_vfnet}, 3) FCOS \cite{Tian2019_fcos}, 4) FoveaBox \cite{Kong2019_foveabox}, and 5) DETR \cite{Carion2020_detr}. Faster RCNN is a two-stage anchor-based detector where the first stage generates region proposals of objects of interest, while the second stage classifies these proposals and regresses the object's bounding box coordinates. As Faster RCNN can generate numerous false positives (FP), Hard Negative Example Mining (HNEM) \cite{Tang2019_ULDOR} is a strategy that can be employed to reduce the number and thereby boost precision and recall values. This method involves sampling negative bounding boxes, which do not overlap with the ground truth and have prediction probability scores that are larger than the true positives, allowing us to iteratively train the model to reduce the incidence of FP.  

FCOS \cite{Tian2019_fcos} is different from the Faster RCNN model; it extracts features from multi-levels of the Feature Pyramid Network (FPN) \cite{Lin_2017_FPN} using shared heads at different levels to regress the bounding box coordinates, and a centerness score to eliminate false predictions that are far away from the center of the target object location. FoveaBox \cite{Kong2019_foveabox} consists of a backbone and a fovea head network, the backbone being a simple FPN yielding feature maps at multi-levels. The fovea head has two branches: the classification branch performs per-pixel classification on the backbone features, and the box regression branch predicts the coordinates of the box at each location that may be potentially covered by an object. 

VFNet combines FCOS (without the centerness branch) with the Adaptive Training Sample Selection (ATSS) module \cite{Zhang2021_vfnet} to accurately select the high quality detections from the pool of candidate predictions. By replacing the classification score with the IoU between the ground truth and the prediction, an IoU-aware classification score (IACS) is obtained that is merged in a novel IoU-aware Varifocal loss, which down-weights the contribution of negative detections to the loss, but up-weights the contribution of positive samples. Inside this setup, a star-shaped bounding box provides context cues to reduce the misalignment between ground truth and the predicted boxes. 

The DEtection Transformer (DETR) \cite{Carion2020_detr} uses a transformer encoder-decoder architecture, wherein the self-attention provided by the transformer removes duplicate predictions. Spatial position encoding is added to the feature maps obtained from the ResNet50 backbone inside the encoder-decoder architecture. Prediction feed-forward network heads decode the output embeddings of object queries independently into the bounding box coordinates and labels. A bipartite set matching loss is used to train the model, however in our implementation, we replaced the original cross entropy classification loss with the focal loss \cite{Lin2017_retinanet} to overcome the class imbalance problem between positive and negative samples, and weighted it with $\lambda_{c}=2$.

\noindent
\textbf{Weighted Boxes Fusion.} At test time, utilizing multiple object detection models or multiple epochs of a single model can yield significantly higher mAP and sensitivity metrics at the cost of a large number of predictions. The predictions consist of the bounding box coordinates of an object and the classification probability. Many strategies, such as Non-Maximal Suppression (NMS) \cite{Bodla_2017_softNMS, Solovyev2021}, exist to reduce these duplicates, but we use Weighted Boxes Fusion \cite{Solovyev2021} to reduce the number of FP during test time.  

\section{Experiments and Results}
\label{sec_Results}

\noindent
\textbf{Data.} Between January 2015 and September 2019, patients who underwent imaging for a lymphoproliferative disease were identified. Initially, 584 T2-weighted abdominal MRI scans were downloaded from the National Institutes of Health (NIH) Picture Archiving and Communication System (PACS). Radiology reports for these studies were also downloaded, and natural language processing \cite{Peng2020} was used to pull the nodal extent and size measurements in these studies. A radiologist performed a quality check on the collected data, and removed incorrect annotations and scans containing only one LN annotation (either LAD or SAD measures). This process yielded 421 T2 scans (n = 421 patients), which were divided into training ($\sim$60\%, 254 scans), validation ($\sim$10\%, 45 scans), and test ($\sim$30\%, 122 scans) splits. The train and validation set consisted of only 2D key slices with 1-2 LN annotated with both the LAD and SAD, while the 3D extent of all LN in the test set were fully labeled. N4 bias normalization \cite{Tustison2010} was first applied to the scans, followed by normalization to [1\%, 99\%] of the voxel intensity range \cite{Kociolek2020}, such that the contrast between bright and dark structures in the slices was boosted. The resulting scans had dimensions in the range from (256 $\sim$ 640) $\times$ (192 $\sim$ 640) $\times$ (18 $\sim$ 60) voxels.

\noindent
\textbf{Implementation.} As radiologists read MRI scans by scrolling through the slices, we replicated their workflow by using 3-slice T2 MRI images where the center slice contained the LN annotated by the radiologist. These 3-channel images were passed to each detection model implemented with the mmDetection framework \cite{Chen2019_mmdet}. Data augmentation was performed: random flips, crops, shifts and rotations in the range of [0, 32] pixels, and [0, 10] degrees etc. ResNet-50 was the backbone for all models used in this paper (pre-trained with MS COCO weights). A grid search was run across the batch size and learning rate parameters to obtain the optimal values; batch size was kept constant at 2 for all models, while the learning rate was 1e-3 for all models (for 12 epochs) except DETR whose learning rate was 1e-5 (for 150 epochs). Each model was executed 5 times, and the best checkpoint from each run with the lowest validation loss was chosen for testing. The ensemble comprised of the lowest validation loss checkpoint for each model across all runs. All experiments were run on a workstation running Ubuntu 18.04LTS with 4 NVIDIA Tesla V100 GPUs. Evaluation was always performed at an IoU threshold of 25\% to be consistent with prior work.

\noindent
\textbf{Results.} From prior work, a clinically acceptable result for LN detection is a sensitivity of 65\% at 4-6 FP per volume \cite{Wang2022_mri,Zhao2020_mri}. Higher detection sensitivity is required as missing any cancerous LN can be troublesome from a screening and triage perspective, however adequate precision is also necessary as too many FP can overwhelm the interpretation of a scan. Our quantitative results are presented in Table \ref{table_LN_detection_results} in terms of mean average precision (mAP) and sensitivity, while qualitative results are shown in Fig. \ref{fig:money}. Our first experiment was to assess the performance of Faster RCNN with and without HNEM respectively. Faster RCNN with HNEM improved LN detection mAP and sensitivity by $\sim$2.2 (48.6\% vs. 46.4\%) and $\sim$3 points (76.9\% vs. 74.1\%) at 4FP respectively over Faster RCNN without HNEM. These results are in accordance with prior work on the utility of Faster RCNN with HNEM in reducing FP \cite{Tang2019_ULDOR}.

In our second experiment, we commissioned the VFNet model to detect LN. VFNet with the ATSS module adequately increased the detection mAP and sensitivity by $\sim$4.7 (51.1\% vs. 46.4\%) and $\sim$4.6 points (78.7\% vs. 74.1\%) at 4FP respectively over Faster RCNN without HNEM. There was a marginal improvement of $\sim$2.5 mAP and $\sim$2 recall by VFNet over Faster RCNN with HNEM. For our third experiment, only the FCOS model in VFNet was used without the ATSS module, and the detection performance drops significantly. The detection mAP dropped by $\sim$11 points (51.1\% vs. 39.6\%), while the sensitivity dropped by $\sim$7 points (78.7\% vs. 72\%). This experiment attests to the utility of the ATSS module within the VFNet model. Note however that these results already show a significant improvement over the current SOTA in LN detection \cite{Wang2022_mri}; the worst performing FCOS model posts a $\sim$7.4 increase in sensitivity at 4FP per volume. 

\begin{table}[!h]
\centering\fontsize{9}{12}\selectfont 
\setlength\aboverulesep{0pt}\setlength\belowrulesep{0pt} 
\setlength{\tabcolsep}{7pt} 
\setcellgapes{3pt}\makegapedcells 
\caption{Detection performance of various detectors and our proposed model ensemble. ``S'' stands for Sensitivity @[0.5, 1, 2, 4, 6, 8, 16] FP. \mbox{--} indicates unavailable metric values.}
\begin{adjustbox}{max width=\textwidth}
\begin{tabular}{@{} c|c|c|c|c|c|c|c|c @{}} 
\toprule
Method                                                          & mAP       & S@0.5     & S@1       & S@2       & S@4       & S@6       & S@8       & S@16 \\
\midrule

Faster RCNN (without HNEM) \cite{Ren2015_fasterrcnn}                 & 46.4      & 42.3      & 54.3      & 65.1      & 74.1      & 77.8      & 78.1      & 78.1 \\
Faster RCNN (with HNEM) \cite{Ren2015_fasterrcnn}                    & 48.6      & 39.4      & 53.2      & 66.5      & 76.9      & 79.5      & 79.5      & 79.5 \\
VFNet \cite{Zhang2021_vfnet}                                    & 51.1      & 45.7      & 56.8      & 67.9      & \textbf{78.7}      & 82.6      & 84.9      & 86.2 \\
FCOS \cite{Tian2019_fcos}                                       & 39.6      & 35.1      & 47.1      & 60.1      & 72        & 76.9      & 79.5      & 81.3 \\
FoveaBox \cite{Kong2019_foveabox}                               & 50.2      & 46        & 57.2      & 68        & 77.4      & 82.1      & 84.1      & 85.6 \\
DETR \cite{Carion2020_detr}                                     & 45.5      & 39.2      & 51        & 61.3      & 74.8      & 79.8      & 81.8      & 82.5 \\

Ensemble (All)                                                  & 49.7      & 44.6      & \textbf{58.4}      & \textbf{69}        & 78.5      & 82.3      & 83.8      & 84.1 \\
Ensemble (One-Stage)                                            & \textbf{52.3}      & \textbf{46.5}      & 58        & 68.9      & \textbf{78.7}      & \textbf{82.7}      & \textbf{85.2}      & \textbf{86.4} \\

\midrule

Wang 2022 \cite{Wang2022_mri} (3D)                              & --        & --        & --        & --        & 64.6      & --        & --        & --   \\
Zhao 2020 \cite{Zhao2020_mri} (3D)                         & 64.5      & --        & --        & --        & --        & --         & 62.6      & --   \\
Mathai 2021 \cite{Mathai2021_mlmi} (2D only)                    & 71.7      & 73.8      & 79.7      & 85.7      & 91.6      & 91.6      & 91.6      & 91.6 \\
Mathai 2021 \cite{Mathai2021_spie} (2D only)                    & 65.4      & 65.4      & 76.1      & 88.1      & 91.7      & 91.7      & 91.7      & 91.7 \\
Debats 2019 \cite{Debats2019_mri} (2D only)                     & --        & --        & --        & --        & --        & --         & 80        & --   \\


\bottomrule
\end{tabular}
\end{adjustbox}
\label{table_LN_detection_results}
\end{table}

The fourth experiment incorporated other one-stage or SOTA models for LN detection. In recent works \cite{Mathai2021_mlmi,Mathai2021_spie}, FoveaBox and DETR have been used to detect LN in T2 MRI scans on a slice level, and the detection metrics for these models were calculated. FoveaBox provided results that were on par with those of VFNet with a $\sim$50.2\% mAP and $\sim$77.4\% recall. The DETR transformer model, however, clocked in a lower detection mAP of $\sim$45.5\% and $\sim$74.8\% sensitivity. For our final experiment, we took all the detection models that had a mAP over 45\% and ensembled them (Faster RCNN with HNEM, VFNet, FoveaBox, and DETR) to be consistent with prior literature \cite{Mathai2021_mlmi} in obtaining results for comparison. Initially, we ensembled all the models together (VFNet, FoveaBox, DETR, and Faster RCNN with HNEM) and noticed that the detection performance did not change significantly; the mAP dropped to $\sim$49.7\% while the sensitivity was $\sim$78.5\%. Then, we ensembled only the one-stage models (VFNet, FoveaBox, DETR) together, and observed the best mAP of $\sim$52.3\% and sensitivity of $\sim$78.7\% at 4FP per volume. 

In prior work specifically focused on LN detection \cite{Debats2019_mri,Zhao2020_mri,Mathai2021_mlmi,Mathai2021_spie,Wang2022_mri}, the models were tested on 2D slices where the LN were annotated with the LAD and SAD, and the detection metrics were computed on only these 2D slices resulting in higher mAP and recall as shown in Table \ref{table_LN_detection_results}. However, this is problematic because it does not reflect the true detection performance, i.e., true positives that have not been marked by the radiologist in the current 2D slice or adjoining slices will be counted as FP when in fact they were detected correctly. Since a desirable automated method should process the whole T2 MRI scan, precision and recall should be associated with the FP per volume and \textit{not} per slice. Therefore, in this work, these models are tested on 122 full 3D scans, where all the LN have been annotated in the test set, and the presented results are more representative of describing the LN existence. From past literature, only two approaches \cite{Zhao2020_mri,Wang2022_mri} have attempted a full 3D volumetric detection of LN in T2 MRI. In \cite{Zhao2020_mri}, a precision of 64.5\% was achieved with a combination of T2 and DWI MRI images, whereas we obtain 52.3\% precision with T2 MRI only. Our one-stage ensemble and VFNet models improve upon the sensitivity of \cite{Zhao2020_mri} by $\sim$16 points (78.7\% sensitivity at 4FP vs. 62.6\% sensitivity at 8FP) and that of \cite{Wang2022_mri} by $\sim$14 points (78.7\% sensitivity at 4FP vs. 64.6\% sensitivity at 5FP). 

\section{Discussion}
\label{sec_Discussion}

Among the various models and ensembles that were presented above, VFNet and the one-stage ensemble (consisting of VFNet, FoveaBox, DETR) met our pre-clinical usage goals; they achieved 78.7\% sensitivity at 4 FP surpassing the required 65\% sensitivity threshold at 4-6 FP \cite{Zhao2020_mri,Wang2022_mri}. These models generated detection results that greatly improved upon prior 3D LN detection work as evidenced by Table \ref{table_LN_detection_results}. From our pipeline, Weighted Boxes Fusion was used as a post-processing step, and this is distinct from the original VFNet model that did not use any post-processing step. As we use multiple models (and checkpoints), we found WBF useful in merging the distinct predictions of various models (or epochs) with the mAP and sensitivity either being maintained or improved. In this paper, DWI is not utilized and its exclusion is a limitation of our work as radiologists in current clinical practice refer to these DWI scans to verify their diagnosis. Adding these DWI scans into the detection pipeline will provide a distinct feature representation in the model layers that can improve the detection results. Furthermore, not providing segmentation supervision in this pipeline is another limitation of our work as supervision from a segmentation branch has been shown to improve detection results \cite{Tang2019_ULDOR}. These are avenues that can be explored in future work. 



\section{Conclusion}
\label{sec_Conclusion}

Detection of lymph nodes in T2 MRI scans aids radiologists in differentiating metastatic nodes from non-metastatic nodes and is useful for staging of lymphoproliferative diseases. In this paper, a CAD pipeline is proposed to identify LN in T2 MRI scans. The VFNet model and the one-stage ensemble provide the best results in terms of detection with $\geq$51\% mAP and 78.7\% sensitivity at 4FP per volume. These results significantly improve upon the state-of-the-art methods \cite{Zhao2020_mri,Wang2022_mri} in lymph node detection with an increase in sensitivity of 16 and 14 points respectively. Our results indicate that the proposed CAD pipeline can be used for detection of lymph nodes in a pre-clinical setting. 

\section{Acknowledgments}
\label{sec_Acknowledgments}

\noindent
\textbf{Funding}: This work was supported by the Intramural Research Programs of the NIH National Library of Medicine and NIH Clinical Center (project number 1Z01 CL040004). We also thank Jaclyn Burge for the helpful comments and suggestions.

\noindent
\textbf{Ethical approval}: All procedures performed in studies involving human participants were in accordance with the ethical standards of the institutional and/or national research committee and the 1964 Helsinki declaration and its later amendments or comparable ethical standards. For this study, informed consent was not required.

\noindent
\textbf{Conflict of Interest}: RMS receives royalties from iCAD, Philips, PingAn, ScanMed, and Translation Holdings. His lab received research support from PingAn. The authors have no additional conflicts of interest to declare.





\clearpage
{\tiny \bibliography{sn-bibliography}}
\bibliographystyle{unsrt}


\end{document}